\documentclass[11pt,a4paper]{article}

\usepackage[utf8]{inputenc}
\usepackage[T1]{fontenc}
\usepackage{lmodern}
\usepackage[english]{babel}
\usepackage{csquotes}
\usepackage{geometry}
\usepackage{setspace}
\usepackage{hyperref}

\title{Conrad Habicht’s 1914 Manuscript on Special Relativity and Einstein’s 1907 Reframing of the 1905 Theory}
\author{Hector Giacomini\\[0.4em]
\small Institut Denis Poisson, Université d'Orléans --- Université de Tours --- CNRS.\\
\small 37200 Tours, France\\
\small \texttt{giacominihector@gmail.com}}
\date{}

\begin{document}

\maketitle

\begin{abstract}
This note examines an apparently unpublished manuscript on special relativity written by Conrad Habicht in 1914 and made available online by the ETH-Bibliothek Zürich in December 2024. To the best of my knowledge, no study of its content has yet been published. Habicht was one of Einstein's closest companions during the Bern years. Between February 1902 and mid-1904 he shared with Einstein many occasions for discussion and companionship in Bern. After leaving the city, he remained in close contact with Einstein through visits, reciprocal stays, and a substantial correspondence extending from the years immediately following 1905 to the eve of the First World War.

The manuscript offers a clear and pedagogical presentation of special relativity. Its historical interest lies in the structure of the exposition and in the memory of the theory that the text preserves. Habicht does not present special relativity as an isolated creation beginning from Einstein's 1905 paper alone. He devotes considerable space to the pre-Einsteinian problem situation: the classical principle of relativity, the ether, Fizeau's experiment, Michelson--Morley, Lorentz's theory, the contraction hypothesis, local time, and the privileged system of the stationary ether. Lorentz is treated as the central figure who brought the electrodynamics of moving bodies to its most acute form before Einstein's intervention.

This note provides a qualitative description of the manuscript, with particular attention to its structure, its treatment of the relation between classical mechanics and electrodynamics, and the respective roles assigned to Lorentz, Michelson--Morley, Einstein, and Minkowski. It also argues that Habicht's exposition stands much closer to Einstein's 1907 review article, \emph{Über das Relativitätsprinzip und die aus demselben gezogenen Folgerungen}, than to the more compressed and self-contained presentation of \emph{Zur Elektrodynamik bewegter Körper} in 1905. The manuscript thus offers an opportunity to examine a striking shift in Einstein's own public presentation of special relativity: from the sparse, principle-based narrative of 1905 to the historically reconstructed and Lorentz-centered exposition of 1907.

The document is therefore a historically significant witness to the early reception and narration of Einstein's theory within the circle of one of his closest Bern companions. Its value lies in showing how, within Einstein's extended milieu, special relativity could be understood not as an isolated act of conceptual creation, but as the principled resolution of a Lorentzian and electrodynamical problem situation.
\end{abstract}

\section{Introduction}

Conrad Habicht occupies a distinctive place in the early biography of Albert Einstein. Born in Schaffhausen in 1876, he studied philosophy at the University of Zurich in 1896--1897, then mathematics and physics at the Zurich Polytechnic, before continuing his studies in Munich and Berlin. He enrolled at the University of Bern on 23 May 1901 to prepare a doctorate in mathematics. His dissertation, devoted to Steiner's circle series, was completed in Bern in 1903 and printed in 1904 \cite{Habicht1904}. After his doctoral work, Habicht pursued a career as a teacher of mathematics and physics: first at the Evangelische Lehranstalt in Schiers from 1904, and later, from 1915 to 1948, at the Kantonsschule in Schaffhausen. He died in Schaffhausen on 23 October 1958 \cite{Seelig1956,CPAE5}.

Einstein had already met Habicht before the Bern years. During his stay in Schaffhausen, from mid-September 1901 to the end of January 1902, Einstein worked as a private tutor in Jakob Nüesch's educational institute. In Schaffhausen he found in Habicht, who was then living in the town, one of the few companions with whom he could share music and intellectual conversation. This early acquaintance became immediately relevant when Einstein moved to Bern at the beginning of February 1902. Habicht was already enrolled at the University of Bern, and the first known letter sent by Einstein after his arrival in the city was a postcard of 4 February 1902 addressed to Habicht. The postcard shows that Habicht was already present in Bern when Einstein arrived, although he was not living in Einstein's immediate neighbourhood \cite{CPAE1,CPAE5}.

Habicht's importance for the history of Einstein's early intellectual milieu derives from his personal and intellectual proximity to Einstein during the Bern years. Together with Maurice Solovine, he became one of the members of the small private discussion circle later known as the Akademie Olympia. Between 1902 and 1904 the three friends met for reading, discussion and companionship. Habicht was also, with Solovine, one of the two witnesses at Einstein's marriage to Mileva Marić in January 1903 \cite{EinsteinSolovine1956,CPAE5}.

\section{The scarcity of close testimonies on the genesis of special relativity}

One of the striking features of the early history of special relativity is the paucity of direct testimony from the people who were closest to Einstein during the Bern years. The problem is not that Einstein lived in intellectual isolation. On the contrary, the available evidence points to a small but real circle of friends, colleagues, and family. The difficulty is that this circle left very few detailed retrospective accounts of Einstein's scientific readings, conversations, and intermediate reflections during the period in which the 1905 relativity paper took shape.

Michele Besso occupies an important place in this context. Einstein explicitly thanked him at the end of the 1905 relativity paper for his loyal assistance and for valuable suggestions \cite{Einstein1905a}. Besso was also Einstein's colleague at the Patent Office from 1904 onward and remained one of his closest friends throughout his life \cite{EinsteinBesso1972,CPAE5}. Yet, despite this exceptional proximity, Besso did not leave a written testimony on his discussions with Einstein in Bern concerning the genesis of special relativity. This silence is historically important. The person whom Einstein himself singled out in the published paper as having assisted him is precisely the person whose private role remains least directly documented by his own hand. The paradox is sharpened by the fate of the Einstein--Besso correspondence itself. The letters that survive do not owe their preservation to an orderly archival transmission, but to a fragile and belated rescue. After Besso's death, Pierre Speziali was able to recover, edit, translate, and annotate the correspondence, eventually publishing the \emph{Correspondance 1903--1955} \cite{EinsteinBesso1972}. The edition brought together more than two hundred letters exchanged between Einstein and Besso and made available a corpus of exceptional importance for the study of Einstein's long intellectual friendship with Besso. Yet this surviving corpus also makes clear the limits of the evidence. It preserves a long friendship, scientific exchanges, philosophical reflections, and later recollections, but it does not provide a direct account of the decisive conversations in Bern in May 1905.

Maurice Solovine presents a different case. He did leave a retrospective account through the introduction to the published volume of Einstein's letters to him. This introduction is valuable. It gives information about the meetings with Einstein and Habicht, the atmosphere of the small Bernese circle, the philosophical readings, and the intellectual habits of the group \cite{EinsteinSolovine1956,Solovine1959}. Yet it does not provide detailed information on Einstein's scientific readings, nor on technical discussions leading to the relativity paper. Solovine's testimony is therefore indispensable for reconstructing the cultural and philosophical environment of the Bern circle, but it remains largely silent on the concrete formation of the 1905 theory.

Conrad Habicht left still less retrospective evidence. No memoir comparable to Solovine's introduction is known. The surviving documentation is mainly composed of Einstein's letters to him, not of Habicht's own account of the conversations, readings, or scientific exchanges that may have taken place \cite{CPAE5}. Habicht therefore appears in the historical record primarily through Einstein's voice. This is particularly frustrating, because Einstein's letters show that Habicht remained an intimate and desired interlocutor, and because his later manuscript on special relativity demonstrates that he was capable of producing a structured exposition of the theory. Yet we do not possess from him a personal recollection of the Bern years or of Einstein's discussions before the completion of the 1905 paper.

The same absence is even more consequential in the case of Mileva Marić. Of all those close to Einstein during the Bern period, she was the person best placed to know his daily readings, habits of work, doubts, and intermediate reflections. She had herself received essentially the same scientific education as Einstein at the Zurich Polytechnic, and she shared with him, at least during the earlier years of their relationship, a common intellectual environment shaped by physics, mathematics, and the ambitions of scientific work \cite{CPAE1}. Yet no known testimony from her provides an account of Einstein's intellectual path toward the final form of special relativity, or more broadly toward the remarkable sequence of papers published in 1905.

This silence matters because it marks one of the most serious gaps in the documentary record. Marić lived with Einstein during the decisive period in which the papers on light quanta, Brownian motion, molecular dimensions, special relativity, and mass--energy equivalence were prepared, discussed, or completed \cite{CPAE2,CPAE5}. She was therefore uniquely situated to observe the practical conditions of his work: the books and journals available in the household, the rhythm of writing, the role of conversations with Besso and Habicht, and the possible continuity between earlier problems and the final published formulations. Whatever one thinks about the extent of her scientific participation, this documentary absence is severe. It deprives historians of the perspective of the person who was closest to Einstein during the decisive months of 1905, and it reinforces the broader difficulty of reconstructing Einstein's intellectual development from the published papers alone, especially since several of those papers provide few explicit bibliographical markers and often leave unstated the readings, discussions, and intermediate sources through which Einstein's arguments may have taken shape.

The result is an unusually asymmetrical situation. The people who could have supplied the most direct information either left no testimony or left only general recollections. The genesis of special relativity must therefore be reconstructed from published texts, later statements by Einstein, scattered letters, and a small number of retrospective accounts \cite{Holton1969,Stachel1982,Stachel1987,Norton2004}.

This documentary scarcity has had an important historiographical consequence. Since the immediate witnesses left few detailed accounts, historians have had to reconstruct Einstein's path to special relativity by emphasizing different kinds of indirect evidence. The result has been a large and diverse literature. Some authors have stressed the philosophical background of the Olympia Academy, especially the readings of Hume, Mach, Pearson, and Poincaré \cite{EinsteinSolovine1956,Solovine1959,Norton2010}. Others have returned to Einstein's later autobiographical recollections, such as the youthful thought experiment of chasing a light beam or the decisive conversation with Besso \cite{Pais1982,Stachel1982}. Others again have taken their point of departure from the internal structure of the 1905 paper itself, especially the magnet-and-conductor example \cite{Einstein1905a,Miller1981,Norton2004}. A further group of studies has reconstructed the electrodynamical problem situation: Maxwell's theory, Lorentz's electron theory, the Michelson--Morley and Fizeau experiments, emission theories of light, and Poincaré's formulation of the principle of relativity \cite{Darrigol2004,Darrigol2005,Janssen2002,Galison2003}. Finally, other studies have placed Einstein's work on relativity against the background of his broader pre-1905 research program, though with different emphases. Abiko has stressed the possible chemico-thermal origins of special relativity, Norton has reconstructed Einstein's path to the 1905 paper in relation to his broader use of principle-theory reasoning, and Renn has situated Einstein's 1905 work within his wider atomistic and statistical research program \cite{Abiko1991,Norton2004,Renn2005}.

The coexistence of these approaches reflects the fragmentary character of the evidence. In the absence of detailed contemporary testimony from Besso, Solovine, Habicht and Mileva Marić, the genesis of special relativity has had to be reconstructed from scattered traces: Einstein's later autobiographical remarks, the known readings of the Olympia Academy, the conceptual structure of the 1905 paper, and the broader printed landscape of electrodynamics around 1900--1905. The plurality of proposed pathways is itself a symptom of the documentary gap.

Habicht's manuscript is therefore significant. It arose from the search for a possible testimony by Habicht on Einstein and relativity. No such memoir or retrospective account seems to have been preserved. Instead, what emerged was not a recollection of conversations, but a sustained manuscript exposition of special relativity \cite{Habicht1914}. This changes the nature of the evidence. The manuscript does not tell us what Habicht remembered of Einstein's work in 1905. It does not report conversations in Bern, nor does it provide a direct narrative of the genesis of the relativity paper. But it does show how a member of Einstein's intimate circle of Bern later organized and presented the theory.

This distinction is essential. Habicht's manuscript cannot be used as direct evidence for the genesis of Einstein's 1905 paper. It is not a diary, not a memoir, and not an eyewitness report of the decisive discussions. Yet it is not an external document either. It was written by someone who had belonged to Einstein's closest circle in Bern, who had received Einstein's private announcement of the 1905 papers, who remained in contact with him for years, and whose scientific and pedagogical competence Einstein himself recognized \cite{CPAE5}. In a documentary situation where Besso, Solovine, Mileva Marić, and Habicht left no direct accounts of the scientific conversations surrounding the relativity paper, this manuscript becomes a valuable indirect trace.

The manuscript helps characterize the intellectual afterlife of those conversations within Einstein's immediate milieu. It shows that, at least for Habicht, special relativity was naturally presented through the conflict between classical mechanics, electrodynamics, the ether, Lorentz's theory, and the experimental problem of ether drift. 

\section{Einstein and Conrad Habicht: a continuing correspondence}

The surviving correspondence shows that, even after the original circle had dispersed, Einstein continued to treat Habicht as an intimate interlocutor and as someone whose presence, judgment, and participation mattered \cite{CPAE5}.

The documentation is, however, strongly asymmetrical. The known and published correspondence consists primarily of Einstein's letters to Habicht. No comparable series of letters from Habicht to Einstein appears to be preserved or published. The relationship is therefore documented mainly through Einstein's voice \cite{CPAE5}. This imbalance must be kept in mind: it prevents a direct reconstruction of Habicht's replies, initiatives, and intellectual position. Yet the surviving letters are still highly informative, precisely because they show how Einstein addressed Habicht and what kind of relationship he assumed with him.

During the years surrounding 1905, Einstein repeatedly tried to draw Habicht back into the Bernese circle. The letters contain invitations to visit, playful reproaches for silence, references to the old ``academy,'' and an insistence that some matters were better discussed face to face than by letter \cite{CPAE5}. The tone is often comic and affectionate, but the content is not merely social. Einstein mixed personal news, domestic arrangements, scientific remarks, and requests for discussion in the same register. This suggests that the intellectual life of the group was not fully captured by written exchange. Much of it must have depended on visits, oral discussion, and informal conversation.

The famous letter of 1905 announcing the four papers is especially important in this context \cite{CPAE5}. The announcement was embedded in the same playful and familiar style that characterizes the rest of the correspondence. Habicht was addressed as someone expected to understand the interest of such a communication.

After 1905 Einstein continued to invite Habicht, to discuss visits, to exchange books, and to involve him and his brother Paul Habicht in technical work \cite{CPAE5}. The project of the so-called \emph{Maschinchen}, an electrostatic device intended for the measurement of small voltages, occupies a notable place in this later correspondence. Einstein's references to experiments, apparatus, repairs, numerical performance, possible publication, and priority show that this was a concrete technical project in which Einstein was involved \cite{Einstein1908,CPAE5}.

Habicht was not only a literary or philosophical companion from Einstein's youth. He belonged to a network in which mathematical, physical, and instrumental questions could circulate. Einstein's later recommendation of Habicht for a teaching position in physics and mathematics also confirms that he regarded Habicht as scientifically and pedagogically competent \cite{CPAE5}. Habicht therefore had the profile of a person capable of understanding and presenting a technically structured exposition of special relativity.

The relationship also continued at the level of domestic and personal sociability. Einstein's letters refer to visits, holiday stays, music, household arrangements, Mileva's presence, and later Habicht's family \cite{CPAE5}. These details show that Habicht remained part of Einstein's private world and that the relationship was not reducible to occasional exchanges of scientific information. It was a durable friendship in which personal, intellectual, and technical elements remained intertwined.

After 1914, however, the dense correspondence of Einstein's Bern and Zurich years becomes much thinner. A late letter from Einstein to Habicht, dated 15 August 1948, nevertheless confirms both the persistence of the old bond and the long silence that had separated the former members of the Bern circle \cite{EinsteinHabicht1948}. Einstein wrote after the death of Habicht's brother, explicitly acknowledging that only such a sad occasion had led him to write again. He recalled the old work with Paul Habicht on the \emph{Influenz-Maschinchen} for measuring small voltages, adding that it had been a beautiful enterprise even if nothing usable had come of it. He also mentioned Solovine, whom he had hosted for several weeks two years earlier. The letter ends with a revealing sentence: despite their mutual silence, Einstein was glad that he and Habicht had not become strangers to one another. The letter thus suggests not a complete rupture, but a long interruption of contact after a period of intense personal and technical intimacy.

\section{The correspondence with Paul Habicht and the technical circle around the \emph{Maschinchen}}

Paul Habicht, Conrad's younger brother, was a trained mechanical engineer and became Einstein's main technical correspondent among the Habicht brothers. The surviving correspondence with Paul offers a useful complement to the much more one-sided documentation of Einstein's relation with Conrad. Whereas Conrad appears chiefly through Einstein's letters to him, a relatively substantial number of Paul's letters to Einstein have been preserved \cite{CPAE5}. These letters often concerned technical devices, experimental difficulties, drawings, calculations, patents, and possible improvements to instruments. They may therefore have been kept because they had practical value.

The correspondence shows that Einstein's relation with the Habicht brothers was not only personal or intellectual, but also technical. From 1907 onward, the \emph{Maschinchen} appears as a shared object of discussion. Einstein wrote jointly to Conrad and Paul, referred to small-scale measurements, asked about the progress of the device, and treated the brothers as close collaborators in a continuing practical project \cite{CPAE5}. Paul's letters, especially from 1908 onward, confirm the same picture from the other side. They are technically inventive, sometimes speculative, and repeatedly seek Einstein's judgment. Paul writes as someone who trusted Einstein not merely as a friend, but as an authority capable of evaluating the feasibility and significance of mechanical and electrical ideas \cite{Einstein1908,CPAE5}.

The range of topics discussed in Paul's letters is broad: electrostatic machines, potential multiplication, measuring devices, contacts, electrical disturbances, patents, and experimental arrangements \cite{CPAE5}. The letters reveal a milieu in which theoretical discussion, mechanical invention, electrical experimentation, and personal friendship overlapped. Einstein was not simply surrounded by abstract conversations in Bern and Zurich; he also moved in a small technical world of instruments, measurements, prototypes, and practical schemes.

Conrad Habicht is less visible in this correspondence, but he is not absent from it. Paul occasionally refers to discussions with Conrad, mathematical problems, shared technical ideas, and the circulation of sketches or information between the brothers \cite{CPAE5}. These scattered references suggest that Conrad belonged to the same practical and intellectual environment, even if the surviving archive gives us only an indirect view of his role. The correspondence with Paul therefore helps to reconstruct the broader Habicht milieu, rather than Paul's activity alone.

After the dense exchanges of the years 1907--1912, the surviving Einstein--Paul Habicht correspondence appears to become very sparse. I have found no evidence of further letters between them during the period 1915--1948, apart from a letter written by Einstein to Paul Habicht from Old Lyme, Connecticut, on 5 August 1935. This late letter is known to me only through the description published in Christie's 2007 catalogue of the Albin Schram Collection \cite{Christies2007}. The catalogue nevertheless describes a historically significant document: a two-page autograph letter in German, accompanied by diagrams of electrical circuits or switching systems, in which Einstein reportedly recalled his youthful work with Paul Habicht on small electrostatic machines, the conversations of the Akademie Olympia, and a political disagreement over Habicht's attitude toward Germany during the First World War.

The letter suggests that the sharp decline of the correspondence after 1914 may not have resulted only from the natural exhaustion of the \emph{Maschinchen} project or from accidents of archival preservation. It may also reflect a real cooling of relations caused by wartime political differences. Although the disagreement is documented through a letter addressed to Paul, the broader loss of contact appears to have affected the Habicht circle as a whole, including Conrad. This makes the letter especially important. Even known only through its auction description, it deserves attention as a rare late testimony linking Einstein's memory of the Habicht circle, their technical collaboration, and the political fractures opened by the First World War \cite{Christies2007}.

This background is important for interpreting Conrad Habicht's later manuscript on special relativity. He belonged to a long-standing circle that had maintained close contact with Einstein through letters, visits, technical projects, and scientific discussion \cite{CPAE5}. The manuscript on relativity was therefore produced within a personal and intellectual environment shaped by many years of direct association with Einstein.

\section{Content and structure of the manuscript}

The relativity manuscript is not signed in the body of the text. Its attribution to Conrad Habicht therefore cannot be inferred from an autograph signature on the manuscript itself. It is, however, explicitly catalogued by the ETH-Bibliothek Zürich as \emph{Über A. Einsteins Relativitätsprinzip. Manuskript ohne Titelüberschrift, ausgearbeitet von Conrad Habicht}, dated 1914 and preserved under the shelf mark Hs 1457:2 \cite{Habicht1914}. The wording of the catalogue describes the manuscript as worked out by him.

This archival attribution can be tested against another document from the same institutional context. The ETH-Bibliothek also preserves an earlier manuscript by Habicht, dated 13 November 1896 and preserved under the shelf mark Hs 1457:1 \cite{Habicht1896}. This document is particularly valuable because it is a sustained handwritten lecture manuscript and ends with the autograph dating and signature: ``Zürich 13. Nov. 96. Conr. Habicht, stud. phil.''
The existence of the signed 1896 manuscript makes the attribution of the 1914 relativity manuscript materially testable.

Habicht's manuscript presents special relativity as the outcome of a long development within classical physics. Its opening pages place Einstein's principle of relativity in continuity with an older idea, already familiar from mechanics: the absence of a privileged state of rest for the formulation of mechanical laws. Habicht begins with the need for a reference system in the description of physical processes. He then introduces the class of systems in which the Galilean principle of inertia holds. From this basis he formulates the classical principle of relativity: if a mechanical process is correctly described in one inertial system, it is described by the same laws in any other system moving uniformly with respect to the first.

The manuscript does not begin with Einstein's 1905 paper, nor with the postulate of the constancy of the velocity of light. It first establishes the deep familiarity of the relativity principle in classical mechanics. Habicht stresses that this principle had become almost second nature to physicists. The problem of special relativity is therefore introduced as a problem created by the apparent failure to extend this mechanical principle to the whole of physics, especially to electrodynamics and optics.

The next part of the manuscript turns to the ether theory of light. Habicht recalls the abandonment of Newton's corpuscular theory in favour of a wave theory and then the incorporation of optics into Maxwellian electrodynamics. Since light propagates through celestial space, the wave theory seemed to require a medium distinct from ordinary matter. This leads to the hypothesis of the ether. Habicht then discusses whether this ether is carried along by matter. Fizeau's experiment occupies an important place in this discussion. The partial dragging of light in moving water is presented as incompatible with a simple theory in which the ether is fully carried along by the moving medium. In Habicht's account, the result leads toward the hypothesis of an ether at rest.

This discussion prepares the central difficulty. If the Earth moves through a stationary ether, terrestrial laboratories should experience an ether wind. Habicht emphasizes the expected scale of such effects. Since the orbital velocity of the Earth is small compared with the velocity of light, ordinary electromagnetic experiments would at most reveal small effects. Yet first-order effects, if present, should still be detectable. Their absence therefore required explanation. At this point the manuscript introduces Lorentz \cite{Lorentz1895}.

Lorentz is presented as the physicist who confronted the problem in its most developed theoretical form. Habicht describes Lorentz's theory as an attempt to bring electrodynamics into agreement with the observed absence of ether-wind effects, not by abandoning the ether, but by supplementing the theory with hypotheses capable of explaining why such effects do not appear. The account is historically important because it gives Lorentz a substantial and explicit role. Lorentz's theory is not treated as a marginal prehistory, but as the immediate theoretical framework within which the problem of moving bodies, the ether, and optical experiments had been posed \cite{Lorentz1895,Lorentz1904}.

The emphasis is even stronger than a general acknowledgement of Lorentz as a precursor. Habicht explicitly situates Lorentz's work at the threshold of Einstein's intervention. After discussing the negative result of Michelson's experiment and the contraction hypothesis, he writes that Lorentz sought to remove the difficulty by means of this hypothesis, adding the chronological indication ``(1904)''. In the same sentence, Habicht states that Einstein, at about the same time, was working on the electrodynamics of moving bodies. This explicit reference to 1904 is significant. It places Lorentz's theory not in a distant prehistory, but in the living theoretical situation immediately preceding Einstein's 1905 paper. Habicht's presentation thus brings Lorentz's 1904 electrodynamics and Einstein's 1905 memoir into close narrative contact \cite{Lorentz1904,Einstein1905a}.

Michelson's experiment---that is, the experiment now usually referred to as Michelson--Morley---receives an especially prominent treatment. Habicht explains its purpose as the detection of second-order effects of the Earth's motion through the ether. He describes the division of a light beam into two perpendicular paths, the recombination of the beams, and the expected displacement of interference fringes upon rotation of the apparatus. The crucial point is qualitative: the experiment should have revealed an effect due to the Earth's motion through the ether, but no such effect was found. Habicht then introduces the Lorentz contraction as the hypothesis devised to account for this negative result. In Lorentz's theory, the contraction is introduced as a real physical change associated with motion through the ether. It allows the null result of Michelson's experiment to be explained while preserving the ether as a privileged frame.

This part of the manuscript gives Lorentz's contraction hypothesis a central narrative function. It is not mentioned merely as a technical device. It is presented as Lorentz's striking answer to the most famous experimental difficulty facing the theory of the stationary ether. The contraction removes the expected fringe displacement, but only at the price of making the ether both physically decisive and experimentally elusive. In Habicht's formulation, the ether becomes at once highly real and inaccessible to direct detection. The result is a theory of considerable power, but also of evident conceptual strain.

Yet Habicht's criticism of Lorentz is not dismissive. The conceptual strain of the theory is presented together with a clear sense of its theoretical importance. Lorentz appears as the physicist who brought the ether theory to its highest degree of sophistication, and who pushed it to the point at which its deepest difficulty became visible. Admiration is therefore compatible with criticism: Habicht identifies the weak points of the Lorentzian construction, but also treats it as a powerful and almost sublime theoretical achievement.

This is the point at which Einstein enters the narrative. Habicht does not present Einstein as beginning from a blank conceptual field. Rather, Einstein is said to have addressed the principled difficulty within the Lorentzian framework itself: the conflict between the relativity principle, familiar from mechanics, and the existence of a privileged ether frame in Lorentzian electrodynamics.

Habicht's wording is especially revealing. According to the manuscript, Einstein showed in his 1905 memoir that it was not necessary to abandon either the relativity principle or the principle of the constancy of the velocity of light. Their compatibility could be restored, but only by subjecting the inherited concepts of space and time to a fundamental critique. Einstein's achievement is described as a principled transformation of a Lorentzian problematic rather than as the external replacement of Lorentz's theory by an unrelated construction. The problem-field is Lorentzian: ether drift, null optical experiments, the transformation of time, contraction, and the privileged system of the stationary ether. Einstein's contribution is to show how this entire structure can be reorganized once simultaneity, time, and length are defined operationally.

The manuscript therefore casts Einstein's contribution as the conceptual reorganization of an already existing problem. The two principles that appear incompatible are the principle of relativity and the constancy of the velocity of light. Habicht explains that Einstein showed that the two need not be abandoned or separated. They can be made compatible, but only if the traditional concepts of space and time are critically revised. In this framework the ether is no longer required. If light has the same velocity in all inertial systems, no single system can be identified as the true rest frame of the light-carrying medium. An ether that would be at rest with respect to two relatively moving inertial systems is no longer representable.

The most extended conceptual discussion in the manuscript concerns time. Habicht turns to the operational definition of time measurement and to the synchronization of distant clocks. He explains that a time assigned to a local event is meaningful when a clock is present at the place of the event. For events occurring at different places, however, the comparison of times presupposes a convention or procedure for synchronizing clocks. The manuscript then presents Einstein's light-signal synchronization procedure. This discussion is central to Habicht's exposition because it shows where the assumption of absolute simultaneity enters classical kinematics and how Einstein's theory removes it.

The derivation of the Lorentz transformation follows from this analysis. Habicht contrasts it with the Galilean transformation, whose final equation expresses the classical postulate of absolute time. He then argues that the propagation of a spherical light wave must be represented in both systems with the same velocity. The Lorentz transformation is thereby introduced as the transformation compatible with the constancy of light speed. The transformation embodies the replacement of absolute time and absolute length by relations depending on the state of motion of the system.

The consequences of the transformation are then discussed. Habicht presents the relativity of simultaneity, the contraction of lengths, and the slowing of moving clocks. An important feature of the exposition is the reinterpretation of the Lorentz contraction. What had appeared in Lorentz's ether theory as a special physical hypothesis is now recovered as a consequence of the new kinematics. The contraction is no longer an ad hoc deformation caused by motion through the ether; it follows from the relativistic relation between measurements made in different inertial systems. This shift is one of the central movements of the manuscript: Lorentz's contraction hypothesis becomes the hinge between the ether theory and Einstein's relativity.

A further feature of the manuscript is its explicit mathematical character. Habicht does not confine himself to a purely verbal exposition: he introduces a number of detailed calculations, in particular in his analysis of Michelson's experiment and in the derivation of the Lorentz transformation. These calculations give the manuscript a pedagogical and technical density that distinguishes it from a merely historical or popular account.

The manuscript also stresses the smallness of relativistic effects at ordinary velocities. This allows Habicht to explain why the new theory does not contradict everyday experience. The theory requires the abandonment of absolute time and absolute length, but the deviations from classical expectations are too small to have been noticed in ordinary circumstances. The text thus combines conceptual radicalness with empirical continuity. Special relativity is presented as a theory that changes the foundations of space and time while preserving the practical validity of classical mechanics in its ordinary domain.

In its final pages the manuscript broadens the discussion. Habicht refers to the invariance of the Maxwell--Lorentz equations under the Lorentz transformation and indicates that the difficulties of the electrodynamics and optics of moving bodies disappear in the new theory. He mentions classical optical tests such as aberration, the Doppler effect, Fizeau's experiment, and Michelson's experiment as receiving simple explanations. The manuscript then turns to Minkowski, whose four-dimensional formulation is presented as a successful mathematical development of Einstein's theory. Space and time are said to form a higher unity, the Minkowskian world in which physical events are ordered \cite{Minkowski1909}.

Finally, Habicht mentions the inertia of energy, discovered by Einstein in 1905 \cite{Einstein1905b}. The equivalence between mass and energy is described as a result of relativistic mechanics and as a unification of the conservation laws of mass and energy. The manuscript closes by briefly connecting this idea with recent experimental results in electron theory and radioactivity.

Habicht does not frame Einstein's relativity through the philosophical influences often associated with Einstein's early epistemology, such as Mach, Hume, or Pearson. Nor does he begin from the magnet-and-conductor thought experiment that opens Einstein's 1905 paper. Instead, the manuscript is organized around the electrodynamical and optical problem of moving bodies: the ether, Fizeau's experiment, Michelson's experiment, Maxwellian electrodynamics, and Lorentz's theory. Lorentz is mentioned fifteen times in the manuscript, whereas Mach, Hume, and Pearson do not appear at all. This distribution confirms that Habicht understood the historical and conceptual entry point into special relativity to lie not in a philosophical genealogy of empiricism, nor in the rhetorical opening of Einstein's 1905 paper, but in the Lorentzian problem-situation out of which Einstein's theory emerged.

The manuscript offers a carefully structured introduction to special relativity. Its most striking feature is the amount of space devoted to the pre-Einsteinian problem situation and, within it, to Lorentz's theory. Lorentz, the ether, Fizeau, and Michelson's experiment are not treated as incidental preliminaries, nor is Lorentz reduced to the role of a mistaken predecessor. They provide the narrative and conceptual framework in which Einstein's intervention becomes intelligible. More precisely, Habicht presents Einstein's 1905 theory as the principled resolution of a Lorentzian problematic: ether drift, null optical experiments, the contraction hypothesis, the transformation of time, and the privileged system of the stationary ether. Habicht's Einstein is the physicist who transforms a conflict already sharpened by electrodynamics, optical experiments, and Lorentz's theory into a new kinematics of space and time.

\section{Einstein's 1907 review article as a likely textual model}

A comparison with Einstein's 1907 review article, \emph{Über das Relativitätsprinzip und die aus demselben gezogenen Folgerungen}, suggests that Habicht's manuscript should not be read merely as a delayed paraphrase of the 1905 paper. Its structure and historical framing are much closer to Einstein's own later exposition of the theory in 1907 \cite{Einstein1907}. This point is important because the 1907 article already represents a significant reorganization of the presentation of special relativity. It is not simply a repetition of \emph{Zur Elektrodynamik bewegter Körper}. It is a retrospective account in which Einstein places the theory within a broader electrodynamical, optical, and experimental context.

The difference from the 1905 paper is striking. In 1905 Einstein opened with the well-known asymmetry between the magnet and the conductor. The observable phenomenon, he emphasized, depends only on the relative motion of the two bodies, whereas the customary electrodynamical description introduces a sharp distinction between the case in which the magnet moves and the case in which the conductor moves \cite{Einstein1905a}. This opening gives the paper a highly compressed and almost self-contained form. Einstein does not begin by reconstructing the history of the ether problem, nor by presenting Lorentz's theory, nor by discussing Michelson--Morley. He begins instead from an internal asymmetry in the existing formulation of electrodynamics and moves rapidly to the two postulates, the definition of simultaneity, the transformation of space-time coordinates, and the kinematical and electrodynamical consequences.

The few historical markers in the 1905 paper remain deliberately general. Einstein refers to the unsuccessful attempts to detect the motion of the Earth relative to the light medium, but he does not name Michelson--Morley or give an account of the experiment. He does not discuss first- and second-order ether-drift effects, does not present the FitzGerald--Lorentz contraction as a historical response to the null result, and does not reconstruct the conceptual architecture of Lorentz's electron theory. The founding paper therefore minimizes the visible genealogy of the problem. Its rhetorical force lies precisely in the way it turns a dense electrodynamical situation into a short chain of principles and operational definitions.

The 1907 review article is organized in a markedly different way. Einstein begins from the classical principle of relativity as it holds in Newtonian mechanics and under the Galilean transformation. He then explains why this principle became problematic once mechanics no longer appeared sufficient as the general foundation of physics and electrodynamics acquired a fundamental role. In this setting, the motion of the Earth relative to the ether becomes a central problem. If the Earth moves through a stationary ether, then optical and electrodynamical effects of this motion should in principle be observable.

Einstein then introduces Lorentz in a substantial way. He recalls that Lorentz had shown, within his electron theory, why first-order effects in the ratio \(v/c\) were not to be expected. This already gives Lorentz a major role in the theoretical preparation of the problem. Yet the difficulty was not thereby removed, because second-order effects could still be expected. It is at this point that Einstein explicitly discusses the Michelson--Morley experiment and the contraction hypothesis of FitzGerald and Lorentz. The negative result of Michelson--Morley, the contraction hypothesis, and the distinction between true time and local time become part of the historical and conceptual staging of the theory.

This shift is not merely a matter of added background. It changes the scenography of special relativity. In 1905, Einstein presents the theory as a compact conceptual solution to an asymmetry in electrodynamics. In 1907, he reconstructs the problem historically and physically: Newtonian mechanics, Galilean relativity, Maxwell--Lorentz electrodynamics, the stationary ether, terrestrial motion through the ether, Lorentz's explanation of the absence of first-order effects, the expectation of second-order effects, the Michelson--Morley experiment, the contraction hypothesis, and finally the Einsteinian reinterpretation. The 1907 article therefore restores much of the Lorentzian and ether-theoretical background that had remained largely implicit or invisible in the 1905 paper.

This is already very close to Habicht's narrative. In both the 1907 review article and Habicht's manuscript, Einstein's theory appears not as an isolated kinematical construction, but as the resolution of a problem sharpened by Lorentzian electrodynamics and by the negative results of ether-drift experiments. Habicht's exposition therefore resembles the 1907 article much more than the spare and programmatic opening of the 1905 paper.

The most important difference between the 1905 and 1907 presentations concerns the role assigned to Lorentz. In 1905, Lorentz is not used as an explicit historical guide to the problem. In 1907, by contrast, Einstein presents Lorentz's theory as having come very close to the principle of relativity. Lorentz had explained the absence of first-order ether-wind effects with remarkable success, and the Lorentz transformations themselves already possessed the formal structure required by the new theory. But Lorentz retained a privileged ether frame and a distinction between the ``true'' time of the ether system and the local time assigned to moving systems. Einstein's own step is then described as the removal of this asymmetry. Lorentz's local time becomes the physical time of the moving system. Length contraction is no longer an additional physical hypothesis caused by motion through the ether; it becomes a kinematical consequence of the new relations between inertial systems.

The treatment of time provides another strong point of difference. In 1905, Einstein introduced clock synchronization directly as the foundation of the new kinematics. The discussion was concise and programmatic \cite{Einstein1905a}. In 1907, the same procedure is embedded in a broader contrast between the old and new conceptions of time. Einstein explains that the time of a system is defined by clocks at rest in that system, synchronized by light signals. This definition is then placed against the Lorentzian distinction between true time and local time. Habicht's manuscript follows precisely this pedagogical path.

The mathematical part of Habicht's manuscript also corresponds more naturally to the 1907 review article than to the 1905 paper. Einstein's 1907 text derives the coordinate-time transformation after a general discussion of equivalent reference systems, homogeneous space and time, and the spherical propagation of a light signal. He then extracts the standard consequences: length contraction, time dilation, the addition theorem for velocities, the impossibility of superluminal signal transmission, the Doppler effect, aberration, Fizeau's experiment, and finally the transformation of the Maxwell--Lorentz equations. Habicht's manuscript is less technical and more pedagogical, but the sequence is recognizably similar.

There are, of course, differences. Habicht writes after Minkowski's 1908--1909 reformulation and therefore includes the four-dimensional interpretation of space and time. He also gives a broader concluding place to the inertia of energy and to the later physical consequences of relativity.

The contrast between the 1907 review and the 1905 paper is instructive. The founding article presented special relativity in a highly compressed and almost self-contained form, with very few explicit historical markers. The 1907 review article restored much of the electrodynamical background: Lorentz's electron theory, the stationary ether, the Michelson--Morley experiment, the contraction hypothesis, and the problem of distinguishing true from local time. Habicht's manuscript stands closer to this second mode of presentation. This makes the manuscript historically valuable as a witness to the way the theory could be understood, within Einstein's close milieu, after Einstein himself had recast it as the resolution of a Lorentzian and ether-theoretical problem.

The broader implication is that the difference between 1905 and 1907 is not only expository. It concerns the public memory and historical staging of the theory. In 1905 Einstein minimized the visible genealogy of the problem. In 1907 he restored a substantial part of the Lorentzian, ether-theoretical, and experimental background that made the theory intelligible as a response to the electrodynamics of moving bodies. Habicht's manuscript belongs to this second regime of presentation. It is historically revealing because it preserves, within Einstein's extended circle, a mode of exposition in which special relativity emerges from the Lorentzian problem-situation rather than from the isolated internal logic of the 1905 paper alone.

Richard Staley has rightly drawn attention to the historiographical character of Einstein's 1907 review article. In \emph{Einstein's Generation}, he treats the emergence of relativity not simply as the result of an isolated theoretical breakthrough, but as the formation of a new disciplinary field in which experimental, electrodynamical, and mathematical traditions were gradually reorganized. Einstein's 1907 review article may be read in this light as an early act of historical framing: it did not merely restate the technical content of the 1905 paper, but placed the relativity principle within a broader electrodynamical and experimental setting, giving a central role to Lorentz's theory and to the Michelson--Morley experiment while presenting Einstein's own formulation as the coherent resolution of earlier difficulties \cite{Staley2008}. In this respect, the 1907 review may be regarded as one of the first histories of relativity written from within the new theory itself. Yet this insight must be qualified in one important respect. The historically reconstructed and Lorentz-centered architecture of the 1907 review did not become Einstein's usual way of narrating special relativity in later years. Its significance lies precisely in its exceptional character: it shows that, two years after the 1905 paper, Einstein could present special relativity as the resolution of a Lorentzian and ether-theoretical problem situation, even though he would not generally continue to tell the story in that form.

The significance of the 1907 review has also been recognized by other authors, but mainly in connection with Michelson--Morley, Fizeau, the equivalence principle, or Einstein's later path toward general relativity \cite{Holton1969,Stachel1982,Stachel1987,Norton2004}. What has not been sufficiently isolated is the change in Einstein's own narrative architecture between 1905 and 1907. What can be stated securely is that the 1907 article gives Lorentz a historical and conceptual visibility that the 1905 paper had not provided. The reasons for this shift remain uncertain. The difference of genre between an original research paper and a review article is certainly important. But the possibility that Einstein had become more sensitive, by 1907, to the need to situate his theory in relation to Lorentz should not be excluded.

\section{Corrections and possible revision layer}

The manuscript contains several corrections and interlinear additions. These corrections are historically noteworthy. They appear at points where the exposition depends on technically precise distinctions: the synchronization of clocks, the comparison of measurements in different inertial systems, the interpretation of Lorentz contraction, the behaviour of moving clocks, and the formulation of relativistic mechanics. Their distribution suggests more than ordinary proofreading. They seem intended to clarify conceptual relations and to strengthen the accuracy of the exposition.

A visual comparison of the handwriting also suggests that at least some of these corrections may not have been written by the same hand as the main text. Habicht himself may have revised the manuscript at a later stage. Nevertheless, the combination of graphic difference and technical placement is significant. It suggests that the manuscript underwent a scientifically informed revision, whether by Habicht himself or by another competent reader.

\section{Conclusion}

The manuscript examined in this note is significant because of its historical position. It does not provide direct evidence about the conversations that preceded Einstein's 1905 paper, but its importance lies elsewhere. It is a sustained pedagogical exposition of special relativity written by Conrad Habicht, one of Einstein's closest companions from the Bern years, and therefore belongs to a very small group of documents connected with Einstein's immediate early milieu.

The manuscript presents special relativity in a historically revealing form. It does not begin with the compressed postulatory structure of Einstein's 1905 article. It begins instead with the classical principle of relativity in mechanics, then moves to optics, Maxwellian electrodynamics, the ether, Fizeau's experiment, the motion of the Earth through a stationary medium, Lorentz's theory, the Michelson--Morley experiment, and the contraction hypothesis. Einstein's intervention is introduced as the conceptual resolution of a difficulty already sharpened by Lorentzian electrodynamics and by the failure to detect ether-drift effects. Lorentz is therefore the physicist whose theory brought the problem to its most developed pre-relativistic form.

Habicht's manuscript presents Einstein as the physicist who reorganizes an existing electrodynamical problem. The distinction between Lorentz's local time and Einsteinian physical time forms the conceptual path by which the reader is led to the Lorentz transformation and to the relativity of simultaneity. In Habicht's presentation, the new theory does not erase its Lorentzian background; it transforms its meaning.

This also explains why the comparison with Einstein's 1907 review article is important. Habicht's manuscript is much closer to the 1907 exposition than to the original 1905 paper. The 1905 article is remarkably austere in its historical framing: it begins with the magnet-and-conductor asymmetry, formulates the two postulates, defines simultaneity, and proceeds directly to the transformation laws and their consequences. It contains no historical reconstruction of Lorentz's theory, no explicit discussion of the Michelson--Morley experiment, and no narrative account of the ether problem. The 1907 review article, by contrast, restores much of this background. It places special relativity within the history of the conflict between the Galilean principle of relativity, Maxwell--Lorentz electrodynamics, the stationary ether, negative optical experiments, and Lorentz's work.

This contrast has a broader methodological consequence. The 1905 article seems to have been written rapidly, in the wake of several other major works of the same year, and its exposition is deliberately compressed. The 1907 article, written with more distance, offers a more reflective reconstruction of the problem situation. It is therefore a particularly useful guide for studying how Einstein could retrospectively present the problem situation out of which the 1905 work emerged.

Habicht's manuscript shows that, within a circle personally close to Einstein, special relativity could be presented through the more historical, Lorentz-centered, and electrodynamical framework that Einstein himself had adopted only two years after the 1905 paper. This manuscript reminds us that the history of special relativity was not transmitted only through the paper of 1905, but also through later acts of exposition, recollection, simplification, and reorganization.

\end{document}